\documentclass[conference]{IEEEtran}
\IEEEoverridecommandlockouts
\usepackage{cite}
\usepackage{amsmath,amssymb,amsfonts}
\usepackage{algorithmic}
\usepackage{graphicx}
\usepackage{textcomp}
\usepackage{xcolor}
\usepackage{graphicx}
\usepackage{amsmath}
\usepackage{multirow}
\usepackage{stfloats}
\usepackage{booktabs}
\usepackage{xcolor, soul}
\sethlcolor{yellow}

\def\BibTeX{{\rm B\kern-.05em{\sc i\kern-.025em b}\kern-.08em
    T\kern-.1667em\lower.7ex\hbox{E}\kern-.125emX}}
\begin{document}

\title{DarkStream: real-time speech anonymization \\with low latency
}

\author{\IEEEauthorblockN{Waris Quamer}
\IEEEauthorblockA{\textit{Dept. of Computer Science and Engineering} \\
\textit{Texas A\&M University}\\
College Station, US \\
quamer.waris@tamu.edu}
\and
\IEEEauthorblockN{Ricardo Gutierrez-Osuna}
\IEEEauthorblockA{\textit{Dept. of Computer Science and Engineering} \\
\textit{Texas A\&M University}\\
College Station, US \\
rgutier@tamu.edu}
}

\maketitle

\begin{abstract}
We propose DarkStream, a streaming speech synthesis model for real-time speaker anonymization. To improve content encoding under strict latency constraints, DarkStream combines a causal waveform encoder, a short lookahead buffer, and transformer-based contextual layers. To further reduce inference time, the model generates waveforms directly via a neural vocoder, thus removing intermediate mel-spectrogram conversions. Finally, DarkStream anonymizes speaker identity by injecting a GAN-generated pseudo-speaker embedding into linguistic features from the content encoder. Evaluations show our model achieves strong anonymization, yielding close to 50\% speaker verification EER (near-chance performance) on the lazy-informed attack scenario, while maintaining acceptable linguistic intelligibility (WER within 9\%). By balancing low-latency, robust privacy, and minimal intelligibility degradation, DarkStream provides a practical solution for privacy-preserving real-time speech communication.
\end{abstract}


\section{Introduction}
Voice recordings contain rich biometric information that reveals not only linguistic content but also personal attributes such as speaker identity, sex, and age, as well as paralinguistics (dialect/accent, emotions). Such sensitive information can be exploited by adversaries for speaker recognition and profiling, raising significant privacy concerns. Consequently, speaker anonymization -- concealing personally identifiable voice attributes while preserving speech intelligibility -- is increasingly crucial with the proliferation of voice-enabled technologies and privacy regulations (e.g. General Data Protection Regulation). Initiatives such as the VoicePrivacy Challenge have established benchmarks for evaluating privacy protection and speech utility in anonymization systems ~\cite{tomashenko2020introducing}. Similarly, government programs (\textit{e.g.,} IARPA’s Anonymous Real-Time Speech program\footnote{https://www.iarpa.gov/research-programs/arts}) emphasize the need for anonymization technologies that can operate under strict real-time constraints. 

However, developing practical low-latency, real-time voice anonymization is highly challenging. Most current approaches \cite{meyer2023anonymizing, pierre22_interspeech} rely on computationally intensive, offline processing, making them unsuitable for interactive applications that demand near instantaneous responses, such as secure voice assistants or anonymous call centers. Even recent streaming models simplify components to reduce latency, often sacrificing linguistic richness and speech quality. This trade-off between latency and quality imposes practical limitations for real-world deployment. For instance, Quamer \textit{et. al.} proposed end-to-end streaming model that employed a lightweight content encoder to achieve $<300 ms$ latency, but using only causal CNNs can limit the fidelity of the anonymized speech \cite{quameranon}.

In this paper, we propose DarkStream, a real-time, low-latency speech anonymization system that addresses these gaps. Our approach builds on existing voice conversion architectures that follow a similar approach, e.g., \cite{quameranon}: we factorize input speech into content and speaker representations and then resynthesize it with a new speaker identity. In contrast to prior streaming work that relied solely on causal CNN encoders, we introduce a causal transformer-based content encoder with limited lookahead to better capture linguistic content without undue delay. Specifically, a short lookahead (on the order of a few tens of milliseconds) feeds a streaming self-attention layer, improving the encoding of phonetic units and coarticulation while bounding latency. To avoid generating any intermediate spectral features, we use a neural vocoder–inspired decoder to produce speech waveforms directly. This one-step waveform synthesis further reduces processing latency and complexity. We evaluate our system’s performance following the VoicePrivacy Challenge 2024 metrics and report objective measures of privacy (speaker verification at equal error rate, EER), utility (automatic speech recognition word error rate, WER, and emotion recognition accuracy, UAR), and runtime latency. The proposed streaming model meets strict real-time requirements on modern GPUs and runs within a few hundred milliseconds latency on CPUs \footnote{https://warisqr007.github.io/demos/darkstream/}. 

\section{Related Work}
Early speaker anonymization methods applied deterministic signal processing techniques, such as shifting formant frequencies using the McAdams coefficient \cite{patino21_interspeech} or modifying pitch contours \cite{tavi2022improving}. While computationally efficient, these approaches are susceptible to reverse engineering due to their predictability. Modern methods instead adopt learning-based voice conversion frameworks. For instance, the VoicePrivacy Challenge baseline extracts speaker embeddings (x-vectors \cite{snyder2018x}), replaces them with embeddings from distant speakers in the embedding space \cite{srivastava20_interspeech}, or generates pseudo-speaker embeddings via generative models \cite{meyer2023anonymizing}. Although effective at confusing speaker recognizers, such methods often degrade naturalness and speaker distinctiveness \cite{yao2024distinctive} and remain vulnerable to advanced de-anonymization attacks.

To address these issues, sophisticated approaches focus on disentangling speaker identity from speech content. Autoencoders combined with adversarial training have been used to remove speaker-specific traits \cite{perero2022x}. Alternatively, controlled manipulation of speaker embeddings via principal component analysis allows targeted shifting of attributes like age or sex \cite{vanrijn22_interspeech}. Factorization and adversarial training techniques have further separated embeddings into distinct attribute spaces. For example, Miao \textit{et al.} \cite{miao2023speaker} employed orthogonal Householder neural networks to remove identity information while preserving speech quality. Additionally, hierarchical generative models provide fine-grained control by decomposing speaker characteristics across multiple levels \cite{hsu2018hierarchical}. Despite their effectiveness, these methods typically require high latency or offline processing, limiting their real-time applicability \cite{quameranon}.

\section{Method}
\begin{figure*}[!]
    \centering
    \includegraphics[scale=0.5]{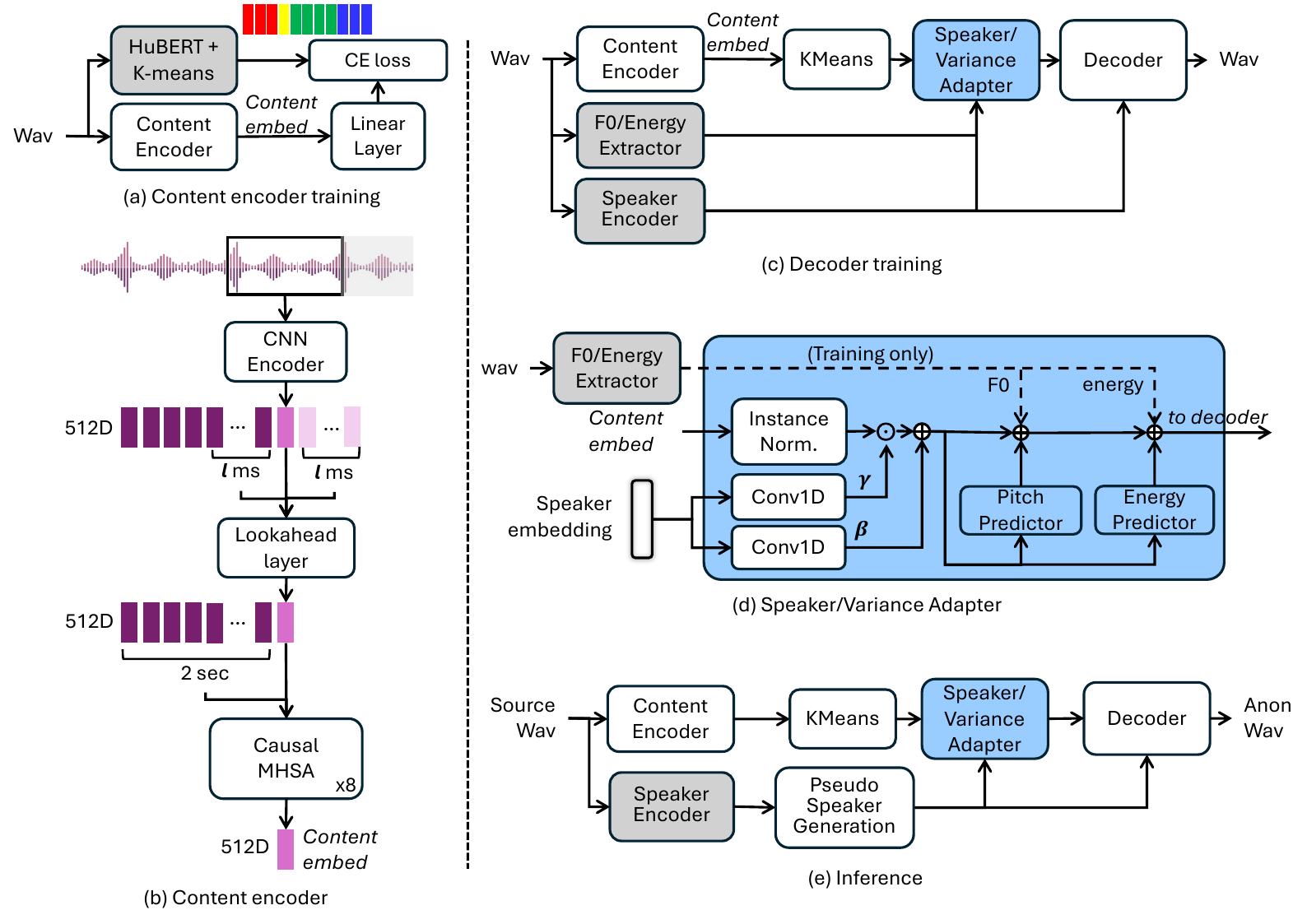}
    \vspace{-13px}
    \caption{Block diagram of the proposed anonymization system. (a) training workflow of the content encoder, (b) content encoder archiecture details, (c) decoder training workflow, (d) speaker/variance adapter archiecture, and (e) inference workflow. }
    \label{fig:block_diagram}
\vspace{-12pt}
\end{figure*}

Our architecture follows similar approach as VC techniques based on deep-learning (DL) and performs speaker anonymization by using a pseudo-speaker embedding as target sampled from a GAN-based generator. As illustrated in Figure \ref{fig:block_diagram}e, our overall framework consists of four main modules: (1) a content encoder that produces linguistic embeddings, (2) a $k$-means bottleneck that suppresses residual speaker cues, (3) a speaker and variance adapter that injects speaker identity (using original speaker embedding as generated through a speaker encoder or pseudo-speaker embeddings) and prosodic attributes, and (4) a decoder that synthesizes waveform given speaker adapted content embedding. Each of these components and their training is detailed below.

\subsection{Content encoder}
The content encoder transforms raw waveform segments into a sequence of 512-dimensional speaker-independent embeddings that represent localized linguistic units. These embeddings are supervised using cross-entropy loss against discrete pseudo-labels obtained by quantizing intermediate HuBERT-base\footnote{https://github.com/facebookresearch/fairseq} (9th layer) representations with $k$-means clustering (with $N=200$ centroids, see Figure \ref{fig:block_diagram}a). In constrast to the streaming anonymization model in \cite{quameranon}, which relied solely on a causal HiFi-inspired encoder, our architecture improves linguistic representation quality by integrating both short-range and long-range contextual cues. As illustrated in Figure \ref{fig:block_diagram}b, it consists of three sequential components: (1) a CNN-encoder that performs frame-level processing in a causal fashion; (2) a lookahead layer that allows limited access to $l$ future audio frames (up to 140–280 ms), enabling the model to capture forward coarticulatory transitions; and (3) a contextual layer composed of stacked causal multi-head self-attention layers, which model long-range dependencies across time.

\textbf{CNN Encoder}. The first component is a convolutional encoder adapted from HiFi-GAN \cite{kong2020hifi} and modified for encoding waveforms in a streaming fashion. Specifically, following \cite{quameranon} we replace all transposed convolutions with strided causal convolutions to downsample the input waveform without violating causality. Each convolutional block processes only past and current audio, ensuring real-time compatibility. In contrast with \cite{quameranon}, however, DarkStream uses a lighter version with one residual block (as opposed to 3 in \cite{quameranon}) and uses downsampling rates of [2,4,5,8] that reduces the temporal resolution of input waveform by a factor of 320 to produce a 512-dimensional latent vector per frame. Each residual block contains convolutions with a fixed kernel size of 5 and dilations of [[1,1],[3,1],[5,1]], capturing short-term acoustic patterns crucial for phonetic discrimination. While this architecture suffices for causal streaming, it lacks anticipatory cues necessary for modeling forward coarticulation and transitions between sounds — which motivates the addition of a lookahead module.

\textbf{Lookahead Layer}. To add to the model’s awareness of near-future context without sacrificing low latency, we introduce a 1D convolutional lookahead layer directly after the CNN encoder. This layer is non-causal, with a receptive field spanning up to 280 ms into the future. This limited future access allows the encoder to model short-term anticipatory effects (e.g., upcoming phoneme transitions) that purely causal models cannot capture. This results in smoother resynthesis and more natural-sounding speech, while keeping the system “near-streaming” with only a minor delay overhead that can be adjusted as needed.

\textbf{Contextual Layer}. Following the lookahead layer, we apply a contextual layer consisting of a stack of $8$ causal multi-head self-attention (MHSA) layers. This layer models long-range dependencies across the encoded sequence, allowing the encoder to form richer linguistic abstractions beyond local patterns. The MHSA operates with causal masking and a fixed past context window (2 secs), which preserves real-time constraints while expanding temporal awareness. Unlike the lookahead layer, which handles short-term future context, the contextual layer allows the encoder to aggregate and propagate relevant information over longer durations, improving consistency and disambiguation over time. We employ a Ring KV cache that stores key and value computations during inference. This cache, initialized with zeros for missing initial context, gradually fills to capacity over 2 seconds.  

% Thus, the model seamlessly transitions from zero context to fully buffered context without performance disruption.

\textbf{Mel-spectrogram based content encoder}. As an alternative front-end, we replace the raw-waveform HiFi based CNN-encoder with a Mel-spectrogram pipeline: the input signal is first converted to a 160-bin Mel spectrogram (16 kHz audio, 1024-sample FFT window, 320-sample hop). This representation retains the spectral structure most relevant to phonetic content while discarding fine-grained phase information, making the subsequent learning problem easier and more robust to channel noise. The Mel frames are then processed by a lightweight ConvNeXt encoder \cite{liu2022convnet} that plays the same role as the HiFi stack, outputting 512-dimensional content embeddings at the original 20 ms frame rate. All downstream blocks--the lookahead convolution and causal self-attention context layer—remain unchanged. In practice, the Mel/ConvNeXt variant offers a favorable quality-versus-latency trade-off and mirrors the success of Mel-based front-ends in modern ASR and TTS systems \cite{liao2024fish}. All ConvNeXt hyperparameters are same as \cite{liu2022convnet} except the number of layers are reduced to [1,1,3,1].

\textbf{$k$-means bottleneck}. Content embeddings are optionally quantized using $k$-means clustering (256 centroids), to remove residual speaker cues. This quantization trades slight synthesis quality for significantly improved speaker anonymization, offering an adjustable privacy-quality balance.

\subsection{Speaker/variance adapter}
We derive speaker identity by concatenating embeddings from two encoders: noise-robust X-vectors \cite{snyder2018x} and context-sensitive ECAPA-TDNN \cite{desplanques2020ecapa}, leveraging their complementary strengths for improved downstream anonymization.

The speaker/variance adapter integrates speaker identity and prosodic features (pitch and energy) into speaker-agnostic content embeddings (\ref{fig:block_diagram}d). First, content embeddings undergo instance normalization to remove residual speaker cues. Two causal 1-D CNNs conditioned on speaker embeddings generate frame-level scale ($\gamma$) and shift ($\beta$) parameters to re-color normalized features via AdaIN/FiLM conditioning \cite{huang2017arbitrary, perez2018film}, aligning them with the target speaker's timbre. Subsequently, speaker-conditioned features pass through separate lightweight blocks predicting fundamental frequency (F0) and energy. Each block consists of a 2-layer causal CNN (kernel=3) with ReLU, layer normalization, dropout, and a point-wise convolution projecting scalar trajectories back into latent space. These blocks are supervised during training by ground-truth F0 and energy. At inference, predictions are directly fed into the feature stream, enabling explicit control of pitch and loudness.

\subsection{Decoder}
The decoder converts a stream of speaker-adapted content embeddings into time-domain audio in two steps steps. The sequence enters the same context layer comprised of a causal MHSA block (similar to the one in the content encoder) with a fixed $2 sec$ look-back window that adds long-range temporal coherence. Then the output sequence is passed to a HiFiGAN-style generator that applies a stack of up-sampling transposed causal convolutions (upsampling rate = [8, 5, 4, 2], residual blocks with fixed kernel size of 5 and dilation as [[1,1], [3,1], [5,1]]) to progressively reconstructs $16 kHz$ audio. The decoder is optimized against objective function and a set of discriminators identical to those in \cite{kumar2023high}: a mel-spec reconstruction loss, multi-scale and multi-period waveform discriminators and multi-resolution spectrogram discriminator.

\subsection{Speaker generation}
We train a Wasserstein GAN with quadratic transport-cost critic (WGAN-QC) similar to \cite{meyer2023anonymizing} to generate pseudo-speaker embeddings. The generator \(G\) maps a 16-dimensional noise vector $z$ through a linear layer into a 192-dimensional latent representation reshaped into a 3×8×8 tensor. This tensor is processed by two ResNet blocks (each consisting of two 3×3 convolutions with residual skip connections), followed by two stages of 
2× upsampling with ResNet blocks, and two final ResNet blocks. The resulting feature map is flattened and linearly projected into a 704-dimensional speaker embedding.

The critic \(D\) reverses this process: it linearly expands a 704-dimensional vector into a 3×8×8 tensor, applies two initial ResNet blocks, then two downsampling stages using 2× average pooling with additional ResNet blocks. Finally, it flattens and maps this representation to a scalar output score.

% To generate pseudo-speaker embedding, we train a similar WGAN-QC model as \cite{meyer2023anonymizing}. Specifically, we train a Wasserstein GAN with a quadratic transport‐cost critic (WGAN-QC) to learn the distribution of valid speaker embeddings and then enforce a minimal angular separation at inference time. We implement generate \(G\) and critic \(D\) using lightweight ResNet‐style networks. The generator \(G\) first maps a 16-dimensional noise vector \(z\) through a linear layer to a 192-dimensional latent, which is reshaped into a \(3\times8\times8\) tensor. It then applies two ResNet blocks (each with two 3×3 convolutions and residual skip), two 2x upsampling + ResNet stages, and two final ResNet blocks. The resulting feature map is flattened and projected via a linear layer to the 704-dimensional speaker embedding.  

% The critic \(D\) reverses this flow: it linearly expands the 704-dim vector into a \(3\times8\times8\) tensor, applies two ResNet blocks, two downsampling + ResNet stages (via \(2\times\) average pooling), and then flattens and linearly maps to a scalar score.  We train the model with the WGAN‐QC losses:

We optimize the networks with WGAN-QC objectives:
\begin{equation}
\footnotesize
\begin{split}
\mathcal{L}_D &= \mathbb{E}_{e\sim p_{\rm real}}[D(e)]
               - \mathbb{E}_z[D(G(z))]
               + 10\,\mathbb{E}_{\hat e}\bigl\|\nabla_{\hat e}D(\hat e)\bigr\|^2,\\
\mathcal{L}_G &= -\mathbb{E}_z[D(G(z))].
\end{split}
\end{equation}

At inference, we sample \(e_{\rm syn}=G(z)\) and reject any sample whose cosine similarity \(\cos(e_{\rm syn},e_{\rm orig})\ge0.65\), ensuring each pseudo-speaker embedding is both realistic and sufficiently dissimilar to the source.

\section{Model training and experimental setup}
We trained the content encoder and decoder on the LibriTTS corpus (train subsets) that provided roughly 600hrs of training data. To train the speaker embedding generator (GAN), we used the CommonVoice (English) \cite{ardila2019common} corpus to cover a diverse range of voice characteristics.  The pretrained speaker encoders were taken from speechbrain \cite{ravanelli2021speechbrain}. 

We evaluate our models on the VoicePrivacy Challenge 2024 evaluation subsets\footnote{https://github.com/Voice-Privacy-Challenge/Voice-Privacy-Challenge-2024}. In particular, we use LibriTTS dev-clean and test-clean sets (for objective intelligibility and privacy metrics), as well as an emotional speech dataset derived from IEMOCAP \cite{busso2008iemocap} to assess how well the anonymization preserves affective content. 
% The IEMOCAP subset includes utterances with varied emotions, and we evaluate emotion recognition accuracy on anonymized vs original speech. 
All the modules were trained using the AdamW optimizer. We set an initial learning rate of $5e^{-4}$ and a batch size of 16 (random 2–4 second clips). The content encoder and decoder were both individually trained for 1.2 million steps on an NVIDIA RTX 3090. When the $k$-means content quantization is applied, we finetuned the decoder with an additional 300k steps: first we freeze the content encoder and run $k$-means on its outputs to establish the 256 centroid codebook. Then, we fine-tune the decoder (and partially the adapter) using quantized content inputs. 

We compare DarkStream against several ablations by systematically varying three factors: (1) the length of the lookahead (LA) buffer, (2) the presence of the contextual layer (CL) with reduced residual blocks in content encoder/decoder, and (3) the input representation (waveform or Mel-spectrogram). For clarity, we note that the waveform-based model without context layer and zero lookahead (Wave at 0ms) is identical to the streaming baseline introduced by Quamer \textit{et al.} \cite{quameranon} and serves as a direct streaming baseline. Additionally, to facilitate a fair comparison across different lookahead settings, we incorporate a lookahead buffer into this baseline. For the Mel-spectrogram front-end without the contextual layer (Mel), we increased the number of residual layers to [3, 3, 9, 3] to match the configuration used by Liao \textit{et al.} \cite{liao2024fish}.

% \vspace{-5pt}
\section{Results}
We evaluate the system along several dimensions: linguistic accuracy, runtime latency, utility (intelligibility and emotion), privacy protection, and subjective quality. Below we present the results for each aspect, with discussions on how the proposed design impacts each metric.

\subsection{Linguistic content preservation}
\label{sec:token_pred}

\begin{table}[t]
\centering
\small
\caption{Top‑1 token‑ID accuracy (\%) as a function of look‑ahead (LA), input representation, and contextual layer (CL).}
\label{tab:la_accuracy}
\vspace{-8px}
\begin{tabular}{@{}lccccc@{}}
\toprule
\textbf{LA (ms)} & \textbf{Wave} & \textbf{Wave\,+\,CL} & \textbf{Mel} & \textbf{Mel\,+\,CL}\\
\midrule
0   & 53.16 & 66.32 & 56.32 & 57.83 \\
20  & 59.82 & 73.37 & 66.09 & 66.78 \\
60  & 61.37 & 77.10 & 73.84 & 74.26 \\
140 & 63.77 & \underline{78.99} & 76.90 & 77.74 \\
280 & \textbf{67.94} &\textbf{ 79.69 }& \textbf{77.75} & \textbf{78.54} \\
\bottomrule
\end{tabular}
\vspace{-10pt}
\end{table}

To quantify how well the system preserves linguistic information, we measure the token (pseudo-phone) prediction accuracy within the content encoder.
Table \ref{tab:la_accuracy} summarizes the top‑1 token‑ID prediction accuracy obtained with our content encoder while varying three factors: the look‑ahead (LA) window, the input representation (raw waveform versus 80‑dim mel‑spectrogram), and the presence of the lightweight contextual layer (CL) that follows the reduced version of causal HiFi stack. Please note that the non-CL wav version uses the original HiFi proposed in \cite{quameranon}. Without LA or CL, the encoder achieves 53.16\% accuracy on waveforms and 56.32\% on Mel‑spectrograms. Adding the CL alone boosts zero‑LA accuracy by +13.2pp for waveforms and +1.5pp for Mels, indicating that the additional temporal receptive field is most beneficial when the front‑end sees the full‑band waveform.

Expanding the LA window yields consistent improvements across all configurations. With no CL, waveform accuracy rises from 53.16\% to 67.94\% when LA grows to 280ms (+27.8\% relative), while the Mel front‑end climbs from 56.32\% to 77.75\% (+38.1\%). When the CL is enabled, gains saturate earlier: waveform+CL reaches 78.99\% at 140ms LA and only nudges to 79.69\% at 280ms, and the Mel variant similarly plateaus at 78.54\%. Because the extra 140ms of look‑ahead yields a marginal $<$1pp absolute improvement at the expense of doubling the buffering delay, we select the 140ms LA with CL as our default streaming configuration. This setting captures $\approx99\%$ of the non‑causal accuracy while keeping the end‑to‑end latency within our 350ms budget.

\subsection{Synthesis latency}
\begin{table}[!t]
\centering
\setlength{\tabcolsep}{1pt}  % <- shrink inter‐column padding
\footnotesize
\caption{Streaming latency (ms) and non‑streaming real‑time factor (RTF) on GPU for a 60ms chunk.  Left: encoder only.  Right: full end‑to‑end pipeline.}
\label{tab:latency}
\vspace{-8px}
\begin{tabular}{@{}lcccccccc@{}}
\toprule
\multirow{2}{*}{\textbf{LA (ms)}} & \multicolumn{4}{c}{\textbf{Encoder only latency (ms)}} & \multicolumn{4}{c}{\textbf{End‑to‑end latency (ms)}}\\
\cmidrule(lr){2-5}\cmidrule(l){6-9}
 & Wave & Wave\,+\,CL & Mel & Mel\,+\,CL & Wave & Wave\,+\,CL & Mel & Mel\,+\,CL\\
\midrule
0   & 82.0 & 69.9 & \textbf{68.2} & 71.7 & 108.2 & \textbf{84.3} & 93.9 & 86.0\\
140 & 200.3 & 189.3 & \textbf{186.2} & 191.6 & 227.1 & \textbf{203.0} & 211.2 & 205.5\\
280 & 320.4 & 309.3 & \textbf{307.2} & 311.4 & 346.9 & \textbf{322.9} & 333.7 & 325.1\\
\midrule
\textbf{RTF} & 0.004 & 0.002 & \textbf{0.001} & 0.002 & 0.010 & \textbf{0.005} & 0.006 & 0.005\\
\bottomrule
\end{tabular}
\vspace{-10pt}
\end{table}

To assess real‑time feasibility we measured (i) encoder‑only streaming latency and (ii) end‑to‑end (e2e) latency, \textit{i.e.,} encoder+decoder on an RTX3090 GPU with a 60ms chunk size. Latency values include both algorithmic delay (primarily look‑ahead buffer and input chunk size) and computational delay per chunk; the complementary real‑time factor (RTF) is obtained by running the same models non‑streamingly.

\textbf{Encoder‑only latency}. Table \ref{tab:latency} shows that without look‑ahead the raw‑waveform front‑end takes 82ms, whereas with the lightweight contextual layer (CL) and reduced HiFi stack reduces computations to 69.9ms (around 15\%) by enabling shallower convolutional depth. Mel‑spectrogram input is even faster at 68.2ms, but here the CL incurs a small 3ms penalty because the lower‑bandwidth representation already enjoys a compact filter bank. Once look‑ahead is introduced, latency grows almost linearly with the buffer size, but the gap between configurations stays within $\pm$11ms. Crucially, all variants run three orders of magnitude faster than real time $(RTF\leq0.004)$.

\textbf{End‑to‑end latency}. The right half of Table \ref{tab:latency} adds the speaker adapter and wav decoder. Zero‑LA latency is still well below 100ms (84.3ms with waveform+CL), and the 140ms look‑ahead setting chosen for accuracy keeps total latency at 203ms, safely inside our 350ms budget. The 280ms setting pushes delay beyond the conversational threshold $\approx320ms$ but the RTF below 0.01 confirms that compute is not the bottleneck—the extra waiting time is purely algorithmic. Across all rows, the waveform+CL configuration offers the best latency/accuracy trade‑off, while Mel‑based models trail by $\leq8ms$. These results demonstrate that the proposed pipeline can operate comfortably in real‑time, even when a 140ms look‑ahead window is used to maximize token‑ID accuracy. On CPU, our wave+CL model achieves RTF of 0.258, roughly 4x faster than real-time.

\subsection{Utility: Intelligibility and Emotion}
\begin{table*}[t]
\centering
\scriptsize
\setlength\tabcolsep{2pt}       % default is ~6pt
\renewcommand\arraystretch{0.9} % default is 1.0
\caption{%
  WER, Emotion-recognition UAR and EER (both lazy-informed and semi-informed scenarios) for different lookahead and input configurations without $k$-means quantization of encoder tokens.}
\label{tab:wer_uar_eer_wout_kmeans}
\vspace{-10px}
\begin{tabular}{@{}l*{16}{c}@{}}
  \toprule
  \multirow{2}{*}{\textbf{LA (ms)}} 
    & \multicolumn{4}{c}{\textbf{WER (mean\,\%, $\downarrow$)}} 
    & \multicolumn{4}{c}{\textbf{UAR (mean\,\%, $\uparrow$)}} 
    & \multicolumn{4}{c}{\textbf{EER (mean\,\%, lazy-informed, $\uparrow$)}} 
    & \multicolumn{4}{c}{\textbf{EER (mean\,\%, semi-informed, $\uparrow$)}} \\
  \cmidrule(lr){2-5} \cmidrule(lr){6-9}
  \cmidrule(lr){10-13} \cmidrule(lr){14-17}
    & Wave & Wave\,+\,CL & Mel & Mel\,+\,CL 
    & Wave & Wave\,+\,CL & Mel & Mel\,+\,CL
    & Wave & Wave\,+\,CL & Mel & Mel\,+\,CL
    & Wave & Wave\,+\,CL & Mel & Mel\,+\,CL \\
  \midrule
  0   & 4.57 & 2.27 & 3.49 & 2.97 
      & 57.00 & 63.88 & 59.70 & 62.62
      & \textbf{36.61} & 20.35 & 24.78 & 15.49
      & 10.20 & 7.02 & 8.58 & 7.51 \\
  140 & 4.17 & \textbf{2.09} & 2.49 & 2.21 
      & 52.35 & 65.31 & 61.96 & \textbf{67.03}
      & 33.57 & 12.10 & 26.82 & 11.92
      & \textbf{10.78} & 6.90 & 8.83 & 8.64 \\
  280 & 4.13 & \textbf{2.09} & 2.36 & 2.13 
      & 55.87 & 64.28 & 60.81 & 65.73
      & 33.67 & 15.51 & 28.06 & 11.96
      & 8.48 & 8.24 & 8.62 & 8.12 \\
  \bottomrule
\end{tabular}
\vspace{-10pt}
\end{table*}

\begin{table*}[t]
  \centering
  % choose one of: \footnotesize, \scriptsize, \tiny
  \scriptsize
  % % narrow columns & rows
  % \setlength\tabcolsep{3pt}       
  % \renewcommand\arraystretch{0.9} 

  \caption{%
    Emotion-recognition UAR and EER (both lazy-informed and semi-informed scenarios) for different lookahead/input configurations with $k$-means quantization of encoder tokens.}
  \label{tab:wer_uar_eer_with_kmeans}
  \vspace{-10px}
  \begin{tabular}{@{}l*{8}{c}@{}}
    \toprule
    \multirow{2}{*}{\textbf{LA (ms)}} 
      & \multicolumn{2}{c}{\textbf{WER (mean\,\%, $\downarrow$)}} 
      & \multicolumn{2}{c}{\textbf{UAR (mean\,\%, $\uparrow$)}}
      & \multicolumn{2}{c}{\textbf{EER (mean\,\%, lazy-informed, $\uparrow$)}} 
      & \multicolumn{2}{c}{\textbf{EER (mean\,\%, semi-informed, $\uparrow$)}} \\
    \cmidrule(lr){2-3}\cmidrule(lr){4-5}
    \cmidrule(lr){6-7}\cmidrule(lr){8-9}
      & Wave+CL & Mel+CL & Wave+CL & Mel+CL
      & Wave+CL & Mel+CL & Wave+CL & Mel+CL \\
    \midrule
    0   & 13.91 & 23.96 & 35.93 & \textbf{34.16}
        & 45.92 & \textbf{47.48} & 19.00 & \textbf{24.19} \\
    140 &  \textbf{9.52} & \textbf{8.75} & 34.49 & 34.73
        &  46.75 & 47.26 & \textbf{22.68} & 21.83 \\
    280 &  9.62 &  8.76 & 35.33 & 34.57
        &  \textbf{47.66} &  46.14 & 22.14 & 22.33 \\
    \bottomrule
  \end{tabular}
\vspace{-15pt}
\end{table*}

Table \ref{tab:wer_uar_eer_wout_kmeans} reports the utility metrics word‑error rate (WER) and unweighted average recall (UAR) when we use the content‑encoder tokens without k-mean quantization. A lower WER means better preservation of linguistic content, while a higher UAR corresponds to better preservation of emotion states. The results follow the same pattern we observed for token‑ID accuracy in Sec. \ref{sec:token_pred}--higher token accuracy translates into lower WER. With no look‑ahead, the CL-based models already halves the WER for waveform input (from 4.57 to 2.27\%) and reduces it by 15\% for mel‑spectrograms (from 3.49 to 2.97\%). Introducing a 140ms buffer brings a further reduction across all configurations, with waveform+CL reaching 2.09\% and mel+CL having a comparable 2.21\%, which are nearly identical to their 280ms counterparts. These diminishing returns mirror the plateau in token‑ID accuracy, reinforcing our choice of the 140ms setting for subsequent experiments. 

UAR broadly follow the trends seen for token accuracy and WER: configurations that yield cleaner linguistic representations also transmit emotion more faithfully. With no look‑ahead, the CL version improves UAR from 57.00 to 63.88\% for waveform input and from 59.70 to 62.62\% for mel‑spectrograms. Introducing a 140ms buffer further raises performance in every setting except “waveform w/o CL,” which drops by 4.6pp, is an outlier. Extending the buffer to 280ms provides no tangible benefit, echoing the diminishing‑returns pattern observed for linguistic metrics.

We repeated WER and UAR experiments with quantized encoder tokens. Results are shown in Table \ref{tab:wer_uar_eer_with_kmeans}. Quantizing tokens with $k$-means introduces an intelligibility penalty (higher WER) yet its impact on downstream utility is somewhat acceptable once a small look-ahead window is available. With no future context, WER roughly doubles compared with the raw-token case, jumping to 13.9\% for waveform+CL and 24.0\% for mel+CL. Expanding the buffer to 140 ms, however, slashes those errors to 9.5\% and 8.8\%, respectively, a 30 to 60\% relative reduction, and pushing the window further to 280 ms yields only a negligible additional gain ($<$0.2 pp). Emotion preservation shows similar pattern: UAR hovers in mid-34\% range once look-ahead is enabled and varies by less than 1pp between 140ms and 280ms, indicating that quantization strips fine-grained prosody without distorting broad affective cues.

\begin{table}[t]
\scriptsize
  \caption{Mean opinion score (MOS) for the proposed system and different ablation for lookahead of 140ms}
  \label{tab:Subjective_MOS}
  \centering
  \vspace{-8px}
  \resizebox{\linewidth}{!}{%
  \begin{tabular}{lllll}
\toprule
    & \textbf{Original} & \textbf{Wav} & \textbf{Wav+CL} & \textbf{Wav+CL+KMeans} \\ \midrule
\textbf{MOS} & 3.90 $\pm$ 0.78  & 3.74  $\pm$ 0.86 & 3.79 $\pm$ 0.85   & 3.22 $\pm$ 0.98 \\ \bottomrule
\end{tabular}}
\vspace{-10pt}
\end{table}
\subsection{Listening tests}
To corroborate the objective results, we conducted a perceptual listening study to validate our findings with human listeners. Specifically, we performed a MOS test with 20 listeners, each rating 20 randomly selected samples from the proposed system and original speech on scale of 1 (poor) to 5 (excellent). Table~\ref{tab:Subjective_MOS} shows that original speech achieved an average MOS of 3.90. Anonymized speech without $k$-means quantization (Wav+CL) obtained an MOS of 3.79, closely matching the original audio quality. Omitting the transformer-based contextual layer (Wav) slightly reduced MOS to 3.74, indicating minor degradation. However, applying $k$-means quantization (Wav+CL+KMeans) led to a notable MOS decrease (3.22), suggesting that quantization introduces perceivable artifacts affecting naturalness. These subjective results align with our objective findings, confirming a trade-off between quantization-induced privacy gains and perceived speech quality.

\subsection{Privacy: speaker anonymization}
To evaluate anonymization strength, we measured equal error rates (EER) under two realistic threat scenarios: a lazy-informed attacker, who knows the anonymization algorithm but lacks clean enrollment data of the target speaker, and a stronger semi-informed attacker, who leverages clean enrollment utterances. Higher EER implies lower linkability (i.e. the system finds it hard to tell if two recordings are the same person). As presented in Table \ref{tab:wer_uar_eer_wout_kmeans}, without token quantization, adding the contextual layer (CL) initially reduces EER, making the anonymized speech slightly easier to identify--likely due to improved linguistic clarity. The same trend can somewhat be observed as the look-ahead buffer grows. 

Introducing vector quantization of encoder tokens via a 256-centroid $k$-means codebook  improves anonymization substantially (see Table \ref{tab:wer_uar_eer_with_kmeans}). Under the lazy-informed scenario, quantization raises EER dramatically to approximately 46-47\%, rendering speaker recognition nearly random and effectively neutralizing this threat model. In the more challenging semi-informed scenario, quantization triples EER to 19-23\%, still far from chance but indicating a significant gain in privacy. Overall, combining a moderate look-ahead window (140 ms), the contextual layer, and $k$-means quantization yields a powerful real-time anonymization framework that effectively balances linguistic quality, latency constraints, and robust speaker identity concealment.

\subsection{VPC24 baselines comparisons}
% \begin{figure}[!th]
%     \centering
%     \includegraphics[scale=0.7]{privacy-ultity_tradeoff.pdf}
%     \caption{Baselines.}
%     \label{fig:baselines}
% \end{figure}
\begin{table}[t]
\scriptsize
  \caption{Baseline EERs from voice privacy challenge 2024}
  \label{tab:baselines_vpc24}
  \vspace{-8px}
  \centering
  % \resizebox{\linewidth}{!}{%
  \begin{tabular}{llllll}
\toprule
    & \textbf{B2} & \textbf{B3} & \textbf{B4} & \textbf{B5a} & \textbf{B5b} \\ \midrule
\textbf{EER} & 5.99        & 26.28       & 31.49       & 22.09        & 34.35 \\
\textbf{WER} & 10.20       & 4.32        & 6.02        & 9.39         & 4.55 \\ \bottomrule
\end{tabular} %}
\vspace{-18pt}
\end{table}

In a final step, we evaluated DarkStream--comprising a waveform or Mel-spectrogram encoder with a contextual layer (CL) and 256-centroid $k$-means quantization, against five baselines from the VoicePrivacy 2024 Challenge \cite{tomashenko2024voiceprivacy}, focusing on the semi-informed attacker scenario (see Table \ref{tab:baselines_vpc24}). This scenario assumes the attacker has access to clean enrollment data, making it a stringent test of privacy preservation.

Baseline B2 employs traditional signal processing techniques, such as formant shifting, and achieves an EER of 5.99\%. B3 utilizes a GAN-based pseudo-speaker generator alongside a cascaded ASR-TTS pipeline, resulting in an EER of 26.28\%. B4 leverages language model-based techniques for anonymization, attaining an EER of 31.49\%. B5a extracts bottleneck features from an ASR model and applies $k$-means clustering with 48 centroids, achieving an EER of 22.09\%. B5b is similar to B5a but uses features from wav2vec2 with the same clustering approach, resulting in the highest EER among the baselines at 34.35\%. DarkStream applies similar pseudo-speaker generation technique as B3 whereas its high-level architecture is closer to B5. \textbf{DarkStream achieves a semi-informed EER of 22.68\%, closely matching the performance of B3 and B5a, despite operating in a streaming fashion with a 140 ms look-ahead, whereas the baselines require a full utterance for processing}. This streaming capability enables lower latency and real-time applicability without compromising privacy.

\section{Discussion}
We introduced Darkstream, a real-time speaker anonymization system that conceals speaker identity while preserving intelligibility and naturalness. The short-lookahead transformer encoder significantly enhanced linguistic fidelity leading to lower WER. Darkstream achieves near-optimal privacy (EER close to 50\%) with minimal degradation in intelligibility (WER).  
% The increased latency (due largely to the 140ms lookahead buffer) substantially enhances linguistic fidelity and intelligibility (WER reductions from ~13\% to ~9\%). Furthermore, under the lazy-informed scenario our k-means quantization module increases the Equal Error Rate (EER) from 36.61\% to 47.66\%, which is substantially closer to
% the optimum EER of 50\%. Moreover, even without the lookahead buffer (at zero lookahead), our model achieves an EER of 45.92\% with
% quantization, at a latency of ~84 ms (20 ms lower than the baseline ~108 ms, [4]), while still delivering better privacy.
At present, however, DarkStream does not explicitly disentangle static speaker traits (\textit{e.g.}, accent, age, sex) from dynamic attributes (\textit{e.g.}, emotion, speaking style), leaving some indirect identity cues potentially intact. Future improvements include explicit disentanglement of static and dynamic attributes, controllable anonymization (e.g., selectable sex or accent), introduction of filler words and speech style modifications to mask habitual patterns, and model optimization for lightweight, CPU-based deployment.

\section*{Acknowledgment}
% To be updated after peer review.
Supported by the Intelligence Advanced Research Projects Activity (IARPA) via Department of Interior/Interior Business Center (DOI/IBC) contract number 140D0424C0066. The U.S. Government is authorized to reproduce and distribute reprints for Governmental purposes notwithstanding any copyright annotation thereon. Disclaimer: The views and conclusions contained herein are those of the authors and should not be interpreted as necessarily representing the official policies or endorsements, either expressed or implied, of IARPA, DOI/IBC, or the U.S. Government.

\bibliographystyle{IEEEtran}
\bibliography{mybib}
\end{document}